\documentstyle[12pt,epsfig]{article}

\newcommand{\be}{\begin{equation}}
\newcommand{\ee}{\end{equation}}
\newcommand{\bea}{\begin{eqnarray}}
\newcommand{\eea}{\end{eqnarray}}
\newcommand{\pert}{perturbative }

\newcommand{\nperts}{non-perturbative }
\newcommand{\DS}{Schwinger-Dyson }

\newcommand{\st}{Slavnov-Taylor }
\def\p{(p\!\cdot\!r)}

\begin{document}

\begin{titlepage}
\begin{flushright}
\large
DTP-99/04\\
January 1999\\
\end{flushright}

\vspace{1.2cm}
\begin{center}
{\large\bf
Perturbative constraints on the Slavnov-Taylor Identity \\
for the Ghost-Gluon Vertex in 
QCD\\}
\vspace*{2cm}
{\bf P. Watson}\footnote{peter.watson@durham.ac.uk} \\
Centre for Particle Theory,
University of Durham \\
Durham DH1 3LE, U.K.\\
\end{center}
\vspace{2.5cm}
\begin{abstract}
\noindent A recent form of the \st identity for the ghost-gluon 
vertex of QCD is compared 
with \pert results.  It is found that this identity, derived assuming 
ghost-ghost scattering can be neglected, is not consistent 
with perturbation theory.  A new identity is derived at the one-loop 
perturbative level.
\end{abstract}
\end{titlepage}

\section{Introduction}
\setcounter{equation}{0}
\baselineskip=6.8mm
\parskip=2mm

\noindent Recently, it has been realised that ghosts may play a key role in the
confining behaviour of quantum chromodynamics (QCD) in covariant gauges.  \DS 
studies of Landau gauge QCD have revealed that the previously neglected ghost 
contributions have an important effect on the infrared behaviour of the 
propagators \cite{lorenz,bloch}.  To arrive at this conclusion, the authors 
were necessarily forced to make certain truncating assumptions, namely 
neglecting ghost-ghost scattering-like contributions in the case of von Smekal 
{\it et al.}~\cite{lorenz} and simply using bare vertices in the case of 
Atkinson and Bloch~\cite{bloch}.

\noindent The problem faced in studying the Schwinger-Dyson equations is that 
the full (non-perturbative) three and four-point functions that occur in the 
equations for the propagators are unknown.  One way to overcome this is to use 
the appropriate \st identities to find the so-called longitudinal 
part\footnote{In simple cases, the longitudinal part can be either the part of 
the vertex that does not vanish under contraction with the vector boson 
momentum or the solution obtained from the \st identity alone.} of the vertex, 
from which it is hoped that sensible physics can be obtained.  One example of 
this is the Ball-Chiu vertex in QED \cite{BC}.  In the case of the ghost-gluon 
vertex, the complete Slavnov-Taylor identity is unknown and identification of 
longitudinal and transverse parts cannot be made.  In \cite{lorenz}, under 
their truncation scheme, an identity was derived and the vertex function 
constructed from the solution.

\noindent In \cite{dav1}, Davydychev {\it et al.} have calculated the ghost two 
and three-point functions perturbatively at one loop order (using dimensional 
regularisation) in arbitrary covariant gauge and dimension. They have gone on, 
in a  more recent paper~\cite{dav2}, to compute the same quantities at two-loop 
order just in the zero-momentum limit. Since the weak-coupling solution to the 
\DS equations is perturbation theory, these results allow a comparison of the 
\pert and \nperts results.

\noindent In this paper, the \pert results for the ghost-gluon vertex are 
substituted into the identity put forward by von Smekal {\it et 
al}~\cite{lorenz} in order to check its validity, the justification being that 
algebraically the full \st identity must be identical to the \pert expression 
at a given order in the coupling.  It is found that the identity is not 
satisfied.  The \pert result at one-loop is then used to derive a form for an 
identity, which is a candidate for {\bf the} \st identity.

\section{Notation and Conventions}
\begin{figure}
\begin{center}
\vspace{-5mm}
\mbox{\epsfig{figure=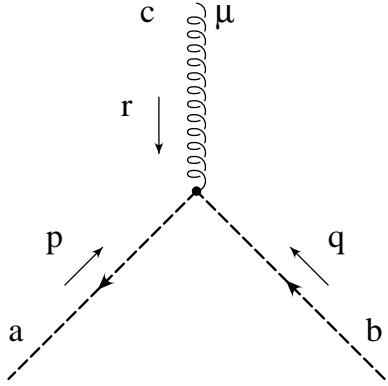,width=5cm}
}
\end{center}
\caption{The ghost-gluon vertex. $a,b,c$ are colour labels, 
$p,q,r$ the momentum routing and $\mu$ the Lorentz index for the gluon.}
\label{fig:vert}
\vspace{5mm}
\end{figure}

\setcounter{equation}{0}
Briefly, the Feynman rules used are written as follows (in Minkowski space).  
The full gluon propagator, with its dressing parameterised by the function 
$J^{-1}$ is
\be
D_{\mu\nu}^{ab}(p)=\delta^{ab} \; \frac{1}{p^{2}} 
\left( t_{\mu\nu}(p) J(-p^{2})^{-1}+\xi\ \frac{p_{\mu}p_{\nu}}{p^{2}}\, ,
\right)
\ee
where $t_{\mu\nu}(p)=g_{\mu \nu}-p_{\mu}p_{\nu}/p^{2}$ is 
the transverse projector and $\xi\!=\!0$ gives the Landau gauge.  The ghost 
propagator with its associated dressing function $G$, is written as
\be
D_{G}^{ab}(p)=\delta^{ab}\;\frac{1}{p^{2}}\;G(-p^{2})
\ee
\noindent The ghost-gluon vertex is defined with all momenta incoming 
and can be 
written as (see Fig.~\ref{fig:vert})
\bea
\widetilde{\Gamma}_{\mu}^{abc}(p,q;r) & \! \equiv \! &
-\imath g \; f^{abc} \; \widetilde{\Gamma}_{\mu}(p,q;r) \nonumber \\
 & \! = \! & -\imath g \; f^{abc} \;
{p}^{\nu} \; \widetilde{\Gamma}_{\nu\mu}(p,q;r)
\eea
\noindent At tree-level, $J=G=1$ and $\widetilde{\Gamma}_{\mu\nu}(p,q;r) 
=g_{\mu\nu}$.  The colour factors are related to the Casimir invariant $C_{A}$ 
by $f^{abc}f^{dbc}=\delta^{ad}C_{A}$.  One-loop expressions for these 
quantities have been calculated~\cite{dav1} in terms of two non-trivial 
functions $\kappa$ and $\varphi$, defined in the following way\footnote{The 
unashamed use of their notation is by virtue of the fact that the present paper 
is based almost entirely on their work.}.  The basic one-loop integral in 
$n=4-2\varepsilon$ dimensions can be written as
\be
I(\nu_{1},\nu_{2},\nu_{3})\equiv\int 
\frac{d^{n}\omega}{[(q-\omega)^{2}]^{\nu_{1}}[(p+\omega)^{2}]^{\nu_{2}}
[\omega^{2}]^{\nu_{2}}}\, ,
\ee
\noindent so that
\be
I(1,1,1)\equiv\imath\pi^{\frac{n}{2}}\eta\varphi,\;\; 
I(1,1,0)\equiv\imath\pi^{\frac{n}{2}}\eta\kappa(r^{2})
\ee
\noindent (similarly for $I(0,1,1)$ and $I(1,0,1)$).  
$\eta$ is a combination of 
$\Gamma$ functions
\be
\eta=\frac{\Gamma^{2}(1-\varepsilon)}{\Gamma(1-2\varepsilon)}\,
\Gamma(1+\varepsilon)\; .\nonumber
\ee
\noindent The function $\kappa$ corresponds to the two-point scalar integral 
and is
\be
\kappa(p)\equiv\kappa_p= 
\frac{1}{\varepsilon(1-2\varepsilon)}(-p^{2})^{-\varepsilon}
\ee
\noindent The two-point function $G$ to one-loop is then
\be
G(-p^{2})\equiv G_{p}=1+\frac{g^{2}\eta}{(4\pi)^{n/2}}\frac{C_{A}}{4} 
\left[(n-1)-\xi(n-3)\right]\kappa(p)\; .
\ee
\noindent The function $\varphi\equiv\varphi(p^{2},q^{2},r^{2})$ is totally 
symmetric and in four dimensions can be written as (see for example 
\cite{dav1},\cite{BigJohnnyC} and references therein)
\be
\varphi=\frac{1}{2p^{2}\sqrt{y^2\!\!-\!x}} \left\{ 
\ln{(x)} 
\ln{\left(\frac{1\!+\!y\!+\!\sqrt{y^2\!\!-\!x}}{1\!+\!y\!-\!\sqrt{y^2\!\!-\!x}} 
\right)} 
\!+\!2{\rm Li}_{2}\!
\left(\frac{x\!+\!y\!-\!\sqrt{y^2\!\!-\!x}}{1\!+\!x\!+\!2y}\right) 
\!-\!2{\rm Li}_{2}\!\left(\frac{x\!+\!y\!+
\!\sqrt{y^2\!\!-\!x}}{1\!+\!x\!+\!2y}\right) 
\right\}
\ee
\noindent where $x={q^{2}}/{p^{2}}, y={p\!\cdot\!q}/{p^{2}}$.  The 
function $\varphi$ therefore encapsulates all the dilogarithmic content 
inherent in any one-loop vertex expression.  The result for the one-loop vertex 
function in the Feynman gauge is (see \cite{dav1})
\bea
\nonumber
\widetilde{\Gamma}_{\mu}(p,q;r) & = & p_{\mu} + \frac{g^{2}\eta}{(4\pi)^{n/2}} 
\frac{C_{A}}{\Delta^{2}} \times \nonumber \\
& & \left\{
p_{\mu}\varphi 
\left[p^{4}r^{2}-\frac{1}{2}p^{2}\p^2+\frac{5}{4}p^{2}\p 
r^{2}+\frac{3}{4}p^{2}r^{4} -\Delta^{2}\left(p^{2}+
\frac{1}{2}\p +\frac{1}{2}r^{2}\right) 
\right] \right. \nonumber \\
& & +r_{\mu}\varphi
\left[ -\frac{1}{2}p^{4}\p 
+\frac{1}{4}p^{4}r^{2}-\frac{3}{2}p^{2}\p^{2} -\frac{3}{4}p^{2}\p r^{2} 
-\frac{1}{2}\Delta^{2}p^{2} \right] \nonumber\\
& & +p_{\mu}\kappa_{p}\left[ \frac{1}{2}\Delta^{2}-\frac{1}{2}p^{2}\p 
-\frac{3}{4}p^{2}r^{2} \right]
+r_{\mu}\kappa_{p}\left[ \frac{1}{2}p^{4}+\frac{3}{4}p^{2}\p \right]\nonumber\\
& & +p_{\mu}\kappa_{q}\left[ \Delta^{2}+\frac{1}{2}p^{2}\p - 
\frac{1}{4}p^{2}r^{2}+\frac{3}{2}\p^{2} +\frac{3}{4}\p r^{2} \right] 
\nonumber\\
& & +r_{\mu}\kappa_{q}\left[ \frac{1}{2}\Delta^{2}-\frac{1}{2}p^{4} 
-\frac{5}{4}p^{2}\p -\frac{3}{4}p^{2}r^{2} \right] \nonumber\\
& & +p_{\mu}\kappa_{r}\left[ -\Delta^{2}+ p^{2}r^{2} -\frac{3}{2}\p^{2} 
-\frac{3}{4}\p r^{2} \right]\nonumber\\
& & \left. +r_{\mu}\kappa_{r}\left[ -\frac{1}{2}\Delta^{2} +\frac{1}{2}p^{2}\p 
+\frac{3}{4}p^{2}r^{2} \right]\right\} 
\eea
\noindent where $\Delta^{2}=p^{2}r^{2}-\p^{2}$.

\baselineskip=7mm
\section{Truncating the Slavnov-Taylor Identity}
\setcounter{equation}{0}
In this section, a short review of the work of von Smekal {\it et 
al}~\cite{lorenz} in deriving a form for the \st identity  is presented.  In 
their scheme, a truncation was made, neglecting all four-point interactions 
including the connected ghost-ghost scattering.  By utilising the BRS 
invariance of the pure Yang-Mills theory, the following identity was found, 
relating the difference of two three-point reducible correlation functions (a 
combination of propagators and contracted ghost-gluon vertices) to the 
four-point function (comprised solely of ghost fields)
\be
\frac{1}{\xi} \langle C^{c}(z)\overline{C}^{b}(y) \partial A^{a}(x)\rangle
- \frac{1}{\xi} \langle C^{c}(z)\overline{C}^{a}(x) \partial A^{b}(y)\rangle = 
-\frac{g}{2}f^{cde} \langle 
C^{d}(z)C^{e}(z) \overline{C}^{a}(x) \overline{C}^{b}(y)\rangle
\label{eq:frog}
\ee
\noindent By then replacing the four-point function with only the disconnected 
ghost propagation terms, they obtain
\be
G_{r}^{-1}\ r^{\mu} \widetilde{\Gamma}_{\mu}(p,q;r)\,+\, 
G_{q}^{-1}\ q^{\mu} \widetilde{\Gamma}_{\mu}(p,r;q)\,+
\,p^{2}\ G_{p}^{-1}\,=\,0\; .
\label{eq:RWI}
\ee
\noindent This equation should be true in all gauges.  One point to notice 
immediately is that as either $q\!\rightarrow\!0$ or $r\!\rightarrow\!0$, one 
simply obtains
\be
p^{\mu} \widetilde{\Gamma}_{\mu}(p,0;-p)\,=\,p^{2}
\ee
\noindent from which one can infer that the vertex in this limit remains bare 
to all orders.  This is known to be true non-perturbatively \cite{mp}.  In the 
wider context it is possible that this behaviour is responsible for the 
similarities of the results for the infrared properties of the gluon propagator 
obtained when using a vertex derived from the above Eq.~(\ref{eq:RWI}) 
and when using simply a bare vertex \cite{bloch}.

\noindent Returning to Eq.~(\ref{eq:RWI}), one finds that this is not true 
perturbatively.  Given that the one-loop expression for the vertex is known, it 
is a straightforward matter to check the identity.  In fact, one does not need 
the full vertex function, but only the part dependent on the integral $\varphi$ 
in the Feynman gauge to see that the identity is not valid except under the 
truncation, since $\varphi$ does not enter the two-point function at this 
order.  Doing this, one obtains
\be
\left[ G_{r}^{-1}\ r^{\mu} \widetilde{\Gamma}_{\mu}(p,q;r) + 
G_{q}^{-1}\ q^{\mu} \widetilde{\Gamma}_{\mu}(p,r;q) + 
p^{2}\ G_{p}^{-1} \right]_{\varphi-dep}\, =\, 
\frac{g^{2}\eta}{(4\pi)^{n/2}}\frac{C_{A}}{4}\  p^{4}\ \varphi\; .
\ee
\noindent In a general covariant gauge, the result is
\begin{eqnarray}
\nonumber
\left[ G_{r}^{-1}\ r^{\mu} \widetilde{\Gamma}_{\mu}(p,q;r) + 
G_{q}^{-1}\ q^{\mu} \widetilde{\Gamma}_{\mu}(p,r;q) + p^{2}\ G_{p}^{-1} 
\right]_{\varphi-dep} \!\! &\!\!=&\nonumber\\[2mm]
\frac{g^{2}\eta}{(4\pi)^{n/2}}\frac{C_{A}}{8}\, \varphi 
\left\{(1+\xi)\ p^{4}\,-\, (1-\xi)\right. \!\!\!\! & & \!\!\!\!\!\! \left. 
(2-n)\ 
p^{2}\ (q\!\cdot\!r)\right\}\,.
\end{eqnarray}
\noindent Thus one can see that the truncation used is not consistent with 
perturbation theory.

\section{Deriving the Slavnov-Taylor Identity from Perturbation Theory}
\setcounter{equation}{0}
In the absence of a full functional derivation of the \st identity, we turn to 
one of the crucial properties of such identities, namely that they are true to 
all orders in perturbation theory.  Eq.~(\ref{eq:frog}) gives us our first clue 
as to the nature of the identity.  The left-hand side is some combination of 
contracted vertices and two-point functions.  The right-hand side is not so 
clear.  In their more familiar guise, \st identities relate three-point 
functions to two-point functions, but here we are faced with a four-point 
correlation function which may or may not be some combination of two and 
three-point functions.  However, it will be shown that there is a simple form 
for the identity based solely on the contracted vertex and the two-point 
dressing function $G$.

\noindent As a first step, consider the $\varphi$ dependence of the quantity 
$r^{\mu}\widetilde{\Gamma}_{\mu}(p,q;r)$ in the Feynman gauge at one-loop.
\bea
\left[ r^{\mu} \widetilde{\Gamma}_{\mu}(p,q;r) \right]_{\varphi-dep}\,=&&  
\frac{g^{2}\eta}{(4\pi)^{n/2}}\frac{C_{A}}{\Delta^{2}}\, \varphi \,
\left\{\frac{1}{2}\ \p^{4} 
+\frac{1}{2}\ \p^3 \ (p^{2}+r^{2})\right.\nonumber\\[2mm]
&-&\left.\frac{1}{4}\ \p^{2}\ 
p^{2} r^{2} -\frac{1}{2}\ \p \ p^{2} r^{2} (p^{2}+r^{2}) -
\frac{1}{4}\ p^{4} r^{4} \right\} .
\eea
\noindent The function $\varphi$ contains all the dilogarithmic content of the 
vertex at this order. This function does not appear in the ghost two-point 
function to this order.  Consequently, it is desirable that it be eliminated 
explicitly.  By inspection, one sees that this part of the contraction is 
symmetric under interchange of $p$ and $r$.  Thus it seems as though a good 
starting point for our identity is
\be
r^{\mu} \widetilde{\Gamma}_{\mu}(p,q;r)-p^{\mu} 
\widetilde{\Gamma}_{\mu}(r,q;p)\,=\,?
\ee

\noindent The right hand side of Eq.~(4.2) is readily deduced using FORM 
\cite{form} for the full vertex function at one-loop \cite{dav1} and one finds 
that
\bea
r^{\mu} \widetilde{\Gamma}_{\mu}(p,q;r)-
p^{\mu} \widetilde{\Gamma}_{\mu}(r,q;p) & 
= & \frac{g^{2}\eta}{(4\pi)^{n/2}}\frac{C_{A}}{8}\, 
\left[(n-1)-\xi(n-3)\right] \times \nonumber\\
& & \hspace{9mm}\left\{\kappa_{p}\ p^{2} - \kappa_{r}\ r^{2} + 
\kappa_{q}\left(r^{2}-p^{2}\right)\right\}\nonumber\\
& = &\frac{1}{2}\ p^{2}\ \left[G_{p}-G_{q}\right]_{one-loop} - 
\frac{1}{2}\ r^{2}\ \left[G_{r}-G_{q}\right]_{one-loop}\, .
\label{eq:PWI}
\eea
\noindent This equation is true in all covariant gauges and dimensions.  That 
this is the only way of eliminating the $\varphi$ dependence of the vertex and 
gives precisely the right structure for the right-hand side to be expressed in 
terms of only the ghost dressing function $G$ leads us to the conclusion that 
this is the one-loop form of the \st identity.  The two quantities on the 
right-hand side each admit four possible \nperts forms, indistinguishable at 
this order
\be
\left[G_{p}-G_{q}\right]_{one-loop} \rightarrow  
\left\{ \begin{array}{c}
       G_{p}\ -\ G_{q} \\[2mm]
       1/G_{q}\ -\ 1/G_{p} \\[2mm]
       1\ -\ {G_{q}}/{G_{p}} \\[2mm]
       {G_{q}}/{G_{p}}\ -\ 1\\[2mm]
       \end{array}
\right.
\ee

\noindent One can clearly see that Eq.~(\ref{eq:PWI}) is unlike the more 
familiar \st identities, since it does not lead to a unique expression for the
so-called longitudinal part of the vertex.  This is disappointing since it is 
usual for the starting point of a vertex ansatz in \DS studies to be based 
around this.

\section{Conclusions}
\setcounter{equation}{0}
A one-loop identity for the ghost-gluon vertex of QCD, valid in all gauges and 
dimensions has been derived from perturbation theory.  It is postulated that 
substituting one of the forms of Eq.~(4.4) into the right hand side of 
Eq.~(4.3) yields the one-loop form of the \st identity.   It is shown to differ
from a previous \nperts identity, obtained using the truncating assumption that 
connected ghost-ghost scattering could be neglected.  The identity does not 
lend itself to a unique definition of the longitudinal part of the vertex. 
Nevertheless, the hope is that  a practical solution of this identity exists, 
which will be useful in \DS studies of the ghost and gluon propagators, and so 
determine their infrared behaviour crucial to confinement.

\newpage

\noindent {\bf Acknowledgements. } The author would particularly like to thank 
M.R.~Pennington for many useful discussions and proof-readings.  
He is grateful to the U.K. Particle Physics and Astronomy Research Council 
(PPARC) for a research studentship.


\begin{thebibliography}{99}
\bibitem{lorenz} L.v. Smekal, A. Hauck, and R. Alkofer, Phys. Rev. Lett. 
{\bf 79} (1997) 3591; Ann. Phys. {\bf 267} (1998) 1; Erratum-ibid. {\bf 269} 
(1998) 182.
\bibitem{bloch} D. Atkinson, and J.C.R. Bloch, Phys. Rev. {\bf D58} (1998) 
094036; Mod.Phys.Lett. {\bf A13} (1998) 1055-1062.
\bibitem{BC} J.S. Ball, and T-W. Chiu, Phys. Rev. {\bf D22} (1980) 2542.
\bibitem{dav1} A.I. Davydychev, P. Osland, and O.V. Tarasov, Phys. Rev. {\bf 
D54} (1996) 4087.
\bibitem{dav2} A.I. Davydychev, P. Osland, and O.V. Tarasov, Phys. Rev. {\bf 
D58} (1998) 036007.
\bibitem{BigJohnnyC} J.M. Campbell, E.W.N. Glover, and D.J. Miller, Nucl. Phys. 
{\bf B498} (1997) 397.
\bibitem{mp} W. Marciano, and H. Pagels, Phys. Rep. {\bf 36C} (1978) 137.
\bibitem{form} J.A.M. Vermaseren, {\it Symbolic Manipulation with FORM} 
(Computer Algebra Nederland, Amsterdam, 1991).
\end{thebibliography}
\end{document}